\pdfoutput=1
\documentclass[aps,groupedaddress,superscriptaddress,twocolumn]{revtex4-1}
\usepackage{graphicx}
\usepackage{dcolumn}
\usepackage{bm}
\usepackage{amssymb,amsfonts,amsmath} 

\newif\ifnumbysec

\def\eqnobysec{\numbysectrue\@addtoreset{equation}{section}}
\newcounter{eqnval}
\newcommand{\rmd}{\mathrm{d}}

\begin{document}

\title{Titration and hysteresis in epigenetic chromatin silencing}
\author{Adel Dayarian}
\affiliation{Kavli Institute for Theoretical Physics,University of California, Santa Barbara, CA}
\author{Anirvan M.~Sengupta}
\affiliation{Department of Physics and Astronomy, Rutgers University, Piscataway, New Jersey}
\affiliation{ BioMaPS Institute for Quantitative Biology, Rutgers University, Piscataway, New Jersey}

\begin{abstract}
Epigenetic mechanisms of silencing via heritable chromatin modiÞcations play a major role in gene regulation and cell fate specification. We consider a model of epigenetic chromatin silencing in budding yeast and study the bifurcation diagram and characterize the bistable and the monostable regimes. The main focus of this paper is to examine how the perturbations altering the activity of histone modifying enzymes affect the epigenetic states. We analyze the implications of having the total number of silencing proteins given by the sum of proteins bound to the nucleosomes and the ones available in the ambient to be constant. This constraint couples different regions of chromatin through the shared reservoir of ambient silencing proteins. We show that the response of the system to perturbations depends dramatically on the titration effect caused by the above constraint. In particular, for a certain range of overall abundance of silencing proteins, the hysteresis loop changes qualitatively with certain jump replaced by continuous merger of different states. In addition, we find a nonmonotonic dependence of gene expression on the rate of histone deacetylation activity of Sir2. We discuss how these qualitative predictions of our model could be compared with experimental studies of the yeast system under anti-silencing drugs.
\end{abstract}

\maketitle

\section{Introduction}
One of interesting biological phenomena  is the possibility for the cells with the same DNA to have different heritable phenotypes.  Such heritable locking of cells into different fates without irreversible change in genetic information is called \textit{epigenetic} phenomenon \cite{allis2007epigenetics}. One of the different mechanisms that can lead to epigenetic effects is transcriptional silencing through chromatin modification. Eukaryotic chromosomes are divided into euchromatin and heterochromatin region, based on the degree of condensation. Euchromatin regions are lightly condensed and genes are accessible to transcription. In contrast, in heterochromatin regions, the chromosome is condensed throughout the mitotic cell cycle and genes are not normally transcribed. Consequently, the formation of heterochromatin is a way of silencing the expression of a number of adjacent genes and stabilizing gene expression patterns in specialized cells \cite{grewal2003heterochromatin}. In order for the cell type to be preserved in cell division, the pattern of heterochromatin and euchromatin regions has to be inherited \cite{alberts-molecular}.  

The first indication for the existence of systematically silenced regions which are inheritable during cell division came from the phenomenon of position effect variegation. Other examples of epigenetic silencing is the \textit{HML} and \textit{HMR} Loci in budding yeast \cite{rusche2003establishment}, and silencing of Hox genes, important in development of body plans, by the Polycomb proteins  \cite{Gilbert:2010fk}. 

Different models have been proposed for silencing in different organisms and even for different regions of the genome in one organism \cite{grewal2003heterochromatin}. However, there is some similarity among some of the proposed mechanisms \cite{moazed2001common}. In general, whether a region of chromosome is in the heterochromatin or euchromatin state depends on the type of modification of histone proteins in the nucleosomes of the corresponding region. Here, we will discuss one of the models which applies to \textit{HML}, \textit{HMR} and telomeric silencing in budding yeast \cite{rusche2003establishment}.

In this model, silencing initiates from a nucleation center which recruits certain proteins including histone modifying enzymes. In the next step, modiÞcation of some of the histone tails of neighboring nucleosomes provides binding sites for the components of silencing complex, which, in turn, modify their neighboring nucleosomes. In such a manner, the silencing region propagates \cite{moazed2001common}. More recent experiments have shown that this picture of linear spreading of silencing from a nucleation center might be only valid for certain loci \cite{radman2011dynamics}. However, the questions that concern us are relevant even if there is only one region where such spreading happens. The key question is what stops the spreading of the silenced region? There are two possible scenarios, suggested by observation from different loci. In some cases, there are explicit boundary elements (e.g. strong gene promoters) stopping the propagation \cite{bi1999uasrpg,donze1999boundaries}. On the other hand, in some other silent regions, experiments perturbing the system  by altering the abundance of pro- and anti-silencing factors lead to graded changes in the extent of silencing  domain \cite{kimura2002chromosomal,suka2002sir2p,radman2011dynamics}.  These observation are consistent with the other possibility that a self-organized stationary state between silenced and active regions is reached, most likely because of the limited supply of the silencing proteins \cite{sedighi2007epigenetic}. We will explore this last possibility in further detail here. 

Biological models of epigenetic silencing suggest bistable dynamics involving positive feedback loops where recruitment of new silencing factors is enhanced by the presence of chromatin bound ones in the neighborhood. Building upon that suggestion, there has been much computational studies of stochastic models of silencing and mean field formulation describing epigenetic states \cite{sedighi2007epigenetic,dodd2007theoretical,david2009inheritance,gils2009quantum,mukhopadhyay2010locus,micheelsen2010theory}. Earlier studies suggest that titration of silencing proteins, caused by the limited supply, has a significant impact on the behavior of the epigenetic silencing system \cite{sedighi2007epigenetic}. The purpose of this paper is to systematically study the different regimes in which such titration causes qualitatively different phenomena, and explore the consequences and predictions of the model.  

One characteristic of nonlinear bistable systems is the hysteresis effects. For example, in a study of the genetic switch in the lac operon system, the two dimensional bifurcation diagram and the corresponding hysteresis effect was explored by changing the abundance of two different molecules which affect the parameters of the system \cite{ozbudak2004multistability}.  In a similar spirit, 
it is possible to study hysteresis in epigenetic silencing by exposing the cells to varying amount of drugs that affect histone modifying enzymes. Such drugs, specially histone deacetylase (HDAC) inhibitors, are already in use as anticancer agents \cite{johnstone2002histone}. In contrast to genetic switches like the lac system \cite{ozbudak2004multistability}, epigenetic chromatin silencing has the additional feature of spreading along chromatin and titrating out silencing factors. This titration effect acts as a negative feedback competing with the positive feedback which gave rise to the bistable behavior in the first place. Therefore, it is important to analyze the interplay between these two phenomena. As we will see, depending upon the abundance of silencing proteins and the size of the regions affected by silencing, one might get very different outcomes in a hysteresis experiment in the silencing system.

\section{Materials and methods}
\begin{figure}[btp]
\includegraphics{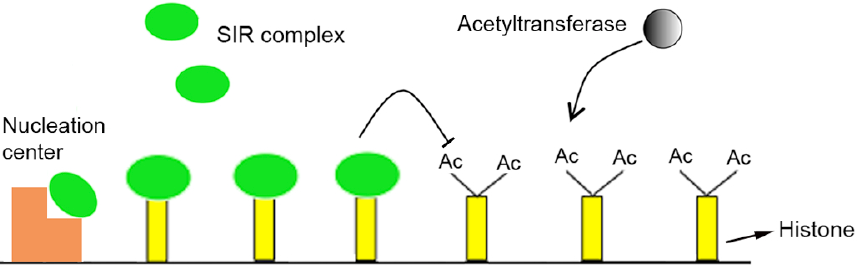}
\caption{A model of epigenetic silencing in budding yeast, \textit{S. cerevisiae}. In this model, bound SIR proteins helps the deacetylation of neighboring histone tails. Recruitment of additional SIR proteins to deacetylated histone tails leads to spreading of silenced region.} 
\label{fig:chromosome_1}
\end{figure}

Much is known about the  silencing proteins in budding yeast and the biochemistry of cooperative interactions between them. The key players are three proteins: Sir2p, Sir3p and Sir4p. These proteins form a complex named SIR (Silenced Information Regulator) complex \cite{rusche2002ordered}. Sir2p is a histone deacetylase which modifies the neighboring histones and provides binding sites for the other proteins in SIR complex, namely, Sir3p and Sir4p \cite{hecht1996spreading,grunstein1998yeast,rusche2002ordered}. 
There are some other proteins which work in an opposing way to the silencing propagation. Particularly, Sas2, a histone acetyltransferase, attaches acetyl groups to certain lysines in histone tails and prevents SIR complex binding \cite{suka2002sir2p,ehrenhofer1997role}. Figure \ref{fig:chromosome_1} shows an schematic presentation of the above model.

In 1989, L. Pillus and J. Rine \cite{pillus1989epigenetic} found that in \textit{sir1} mutants (where the nucleation effect is defective, if not absent), a population of yeast cells is divided into two distinguishable groups. In one group, \textit{HML} locus is silenced similar to normal cells. In the other group, those loci are active and cells would not mate like a normal haploid. Both of the epigenetic states (silenced vs active) are quite stable and are inherited most of the time during cell division. This observation suggests that the system can be thought of as being in a bistable regime, where two stable states can exist under the same external condition. In the model  studied below, this bistability is due to competition between opposing forces and cooperatively in the binding of SIR proteins. However, the conclusions drawn later in this study are not affected by the explicit mechanism of cooperatively.

\subsection{Stochastic equations describing silencing mechanism}
\begin{figure}[btp]
\includegraphics{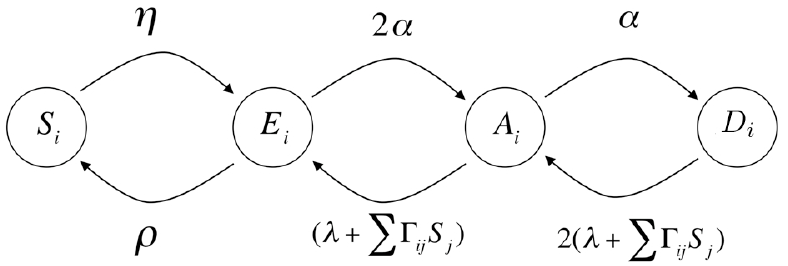}
\caption{Four possible states for each nucleosome. Chromosome is modeled as a 1 dimensional lattice, where each site $i$ corresponds to one nucleosome. Site $i$ can be either in silenced ($S_i$), unbound ($E_i$), monoacetylated ($A_i$) or double-acetylated ($D_i$) state. The deacetylation rate depends on the silencing state of neighboring nucleosomes (the term $\Gamma_{ij} S_j$).} 
\label{fig:stochastic}
\end{figure}

One can think of each chromosome as a 1 dimensional lattice, where each site corresponds to a nucleosome. Each site $i$ can be in one of four possible states: 
\begin{itemize}
  \item Bound by silencing proteins with probability $S_i$
  \item Not bound by any proteins with probability $E_i$ 
  \item Bound by one acetyl group with probability $A_i$
  \item Bound by two acetyl groups with probability $D_i$
\end{itemize}
Figure \ref{fig:stochastic} shows these possible states and the transition rates between them. The rate of SIR binding, which is a function of the ambient SIR concentration, is denoted by $\rho$. Free Sir2p, Sir3p and Sir4p proteins in the environment do not form SIR complex. Instead, they form the complex when they are attached to a nucleosome. In the case where each protein is in low abundance, $\rho$ is proportional to the product of the three concentrations for Sir2p, Sir3p and Sir4p. For our analysis, we will not need to know the exact form of dependence of $\rho$. We will just keep it as an effective parameter, monotonically increase with the concentration of each of SIR proteins. 

The histone acetylation rate, caused by Sas2 activity, is represented by $\alpha$. The rate at which SIR complex falls off the nucleosomes is shown by $\eta$. Also, the basal rate at which acetyl groups are removed from the nucleosomes is denoted by $\lambda$. The deacetylation rate increases if adjacent sites are in the silenced state. This increase is given by the term $\Gamma_{ij} S_j$, where $\Gamma_{ij}$ is a function of $|i-j|$ and drops significantly as this separation increases. All the above parameters may be position and/or time dependent. However, for the sake of brevity, this dependence is not explicitly written. 

We have included a double acetylation state, based on the fact that each nucleosome has two H4K16 sites. It should be mentioned that there is no experimental evidence for cooperativity in the acetylation of these two sites. The reason we include a double acetylation state is to get enough nonlinearity in the system to achieve bistability (see below). This nonlinearity could be provided by other players and degrees of freedom not included in our  model. Most of our conclusion regarding titration and bistability is independent of the particular mechanism by which this nonlinearity is introduced. 

In our model, we do not explicitly incorporate the cell division. Instead, we use a uniform rate, denoted by $\lambda$, that partly models the dilution of silencer-bound histones caused the by the cell division. In another study \cite{david2009inheritance}, one of the authors has addressed this question in more details. Based on that study, we believe that explicitly modeling the cell-cycle does not change many of the essential conclusions regarding titration.

Under the above assumptions, one gets the following chemical kinetics equations:  
\begin{eqnarray}\label{silencing_discrete} 
\frac{\rmd S_i}{\rmd t} & = & \rho  E_i - \eta S_i,   \\
\frac{\rmd E_i}{\rmd t}  &= & ( \eta S_i   -\rho E_i) +  ((\lambda+\Gamma_{ij} S_j) A_i   -2\alpha  E_i ), \nonumber \\
\frac{\rmd A_i}{\rmd t}  &=  & 2\alpha  \hspace{1mm}  E_i +2 (\lambda+\Gamma_{ij} S_j) D_i - (\alpha+\lambda+\Gamma_{ij} S_j) A_i,  \nonumber  \\
\frac{\rmd D_i}{\rmd t}  &=  & \alpha  A_i -2 (\lambda+\Gamma_{ij} S_j) D_i . \nonumber
\end{eqnarray}
We will consider both the uniform solutions and non-uniform solutions in the continuum limit of  these equations.

\subsubsection{Uniform solutions}
We consider uniform steady state solutions for the set of equations \ref{silencing_discrete}, namely, we drop the subscript $i$ and put the left hand sides equal to zero. Let us define $\gamma=\Sigma_j\Gamma_{ij} $. Using the above notation, the uniform solutions of the set of equations \ref{silencing_discrete} has to satisfy:
\begin{eqnarray} \label{silencing_uniform}
0 & = & \rho  E - \eta S     \\
0 & = &  ( \eta S   -\rho E) +  ((\lambda+\gamma S) A   -2\alpha  E)  \nonumber  \\   
0 & = & 2\alpha  \hspace{1mm}  E +2 (\lambda+\gamma S) D - (\alpha+\lambda+\gamma S) A   \nonumber \\
0 & = & \alpha  A -2 (\lambda+\gamma S) D  \nonumber
\end{eqnarray} 
One can divide the above equations by $\lambda$ and redefine the rest of the parameters in units of  $\lambda$: 
\begin{eqnarray}  
\frac{ \rho}{\eta} \rightarrow \rho;  \qquad \frac{ \alpha}{\lambda} \rightarrow \alpha; \qquad  \frac{\gamma}{\eta} \rightarrow \gamma\nonumber . 
\end{eqnarray}
Note that these redefined parameters are dimensionless since they represent the ratios of the original parameters to $\lambda$. Under the simplest of assumptions, $\lambda$ is equal to the rate of cell division, since the new histones that are introduced at the cell division dilute out the silencer-bound ones. In that case, $\lambda \sim 1/$hour. The results of Cheng and Gartenberg \cite{cheng2000yeast} are consistent with this assumption. In practice, some loci have higher rates of histone turnover \cite{dion2007dynamics}, suggesting the rate could be faster in a locus-specific way.

For each set of parameters $\alpha$, $\rho$ and $\gamma$, we would like to be able to characterize how many solutions exist. The derivation of the solutions for the above equations is presented in \ref{APDUS}. It turns out, depending on the value of the parameters, there can be one or two stable solutions. The bifurcation diagram helps to visualize different parameter values which give rise to these two different regimes of monostability and bistability. 

\subsubsection{Bifurcation diagram}
As explained in \ref{APDUS}, by changing the parameters continuously, one can switch between the two regimes of monostability and bistability.  At the point where this transition happens, the following equation has to be satisfied:
\begin{eqnarray}\label{bif_sol_3_maintext}
\gamma & = & \left(S-2S^2-\sqrt{\frac{4(1-S)S^3}{\rho}}\right)^{-1} \  , \\ 
\label{bif_sol_4_maintext}
\alpha & = & \frac{\left(2(1-S)-\sqrt{4\rho^{-1}(1-S)S}\right) \left(\sqrt{\rho S(1-S)} - S  \right) }{S\left(1-2S-\sqrt{4\rho^{-1}(1-S)S}\right)}. \nonumber \\
\end{eqnarray}
The variable $S$ can be replaced by any value between 0 and 1, as long as parameters remain positive real numbers. These conditions can also be rewritten in the following format:
\begin{eqnarray}\label{bif_sol_5_maintext}
\gamma & = & \frac{(\alpha-2(1+\alpha S))}{2S} \pm \sqrt{\left[\frac{(\alpha-2(1+\alpha S))}{2S}\right]^2-\frac{(1+\alpha)}{S^2}} , \nonumber \\ \\ \label{bif_sol_6_maintext}
\rho & = &  \frac{4(1-S)S^3}{\left(S-2S^2-\gamma^{-1}\right)^2} .
\end{eqnarray}

The derivation of the above conditions are presented in \ref{APDBD}.  Equations \ref{bif_sol_3_maintext} and \ref{bif_sol_4_maintext} can be used to draw a plane in the three dimensional  $\rho$ - $\alpha$ - $\gamma$ coordinates. In fact, equations \ref{bif_sol_5_maintext} and \ref{bif_sol_6_maintext} give exactly the same plane in the 3-dimensional coordinates. This plane separate the the two regimes of monostability and bistability. It is  convenient to draw the intersections of this plane with, for example, the constant $\rho$ or the constant $\alpha$ surface. To get the former one, we should keep $\rho$ in equations \ref{bif_sol_3_maintext} and \ref{bif_sol_4_maintext} constant. Instead, for the later case, we should keep $\alpha$ in equations \ref{bif_sol_5_maintext} and \ref{bif_sol_6_maintext} constant. In the next section, we find it convenient to work with equations \ref{bif_sol_5_maintext} and \ref{bif_sol_6_maintext}. However, for now, we stick with equations \ref{bif_sol_3_maintext} and \ref{bif_sol_4_maintext}.

\begin{figure}[btp]
\includegraphics{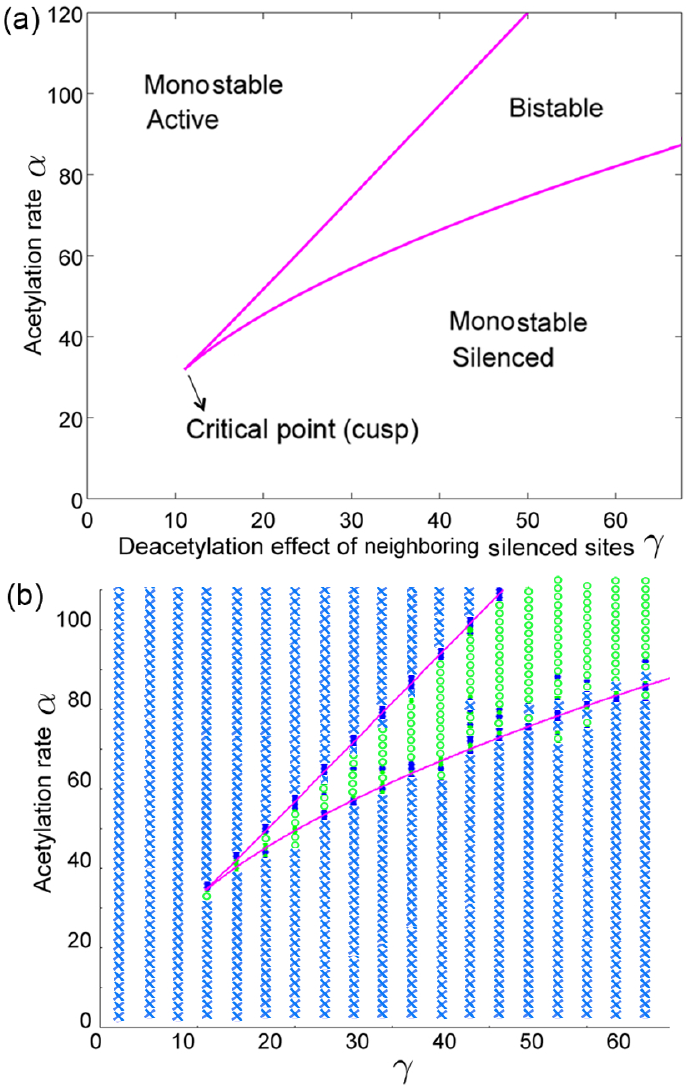}
\caption{Bifurcation diagram in the constant $\rho$ plane. (a) The diagram for $\rho=30$. The magenta line is given by equations \ref{bif_sol_3_maintext} and \ref{bif_sol_4_maintext}.  (b) Result of stochastic simulations. The blue and green points show the two monostable and bistable regions, respectively. The system is simulated starting from two different initial conditions (high and low $S$). Monostable and bistable regimes can be differentiated based on whether the two initial conditions lead to one (blue cross) or two (green circles) different final states.} 
\label{fig:alpha_gamma}
\end{figure}

Figure \ref{fig:alpha_gamma} shows the bifurcation diagram in the $\alpha$ - $\gamma$ plane (constant $\rho$).  If $\alpha$ (acetylation rate) is relatively low but $\gamma$ (deacetylation effect of neighboring silenced sites) is high, only one solution is possible. In this case, most sites are in the silenced state (see \ref{APDBD}). The opposite happens for high $\alpha$ and  low $\gamma$ values. There also exist an intermediate range of values for $\alpha$  and $\gamma$ for which the system is in the bistable regime. We will denote these two stable solutions by $S_h$ and $S_l$, referring to high and low silencing value. At the cusp is the critical point, where all the  solutions merge together (equation \ref{bif_sol_7}). 

For the two stable solutions at each point in the bistable regime, we would like to know which one is more stable. Another interesting question has to do with the fact that we are dealing with an spatially extended system. One may wonder whether it is possible to have different parts of the system to be in different states (silenced vs active) with an stable boundary between them. The subject of the next section is addressing such issues.  

\subsection{Non-uniform solutions, coexistence of domains}
In addition to uniform solutions discussed before, we would like to explore the possibility of having non-uniform spatial solutions for the parameter sets located in the bistable regime. In other words, we are looking for a solution which starts from one of the stable solutions (e.g. active state) and ends in the other one (e.g. silenced state).  As we mentioned in the previous section, heterochromatin and euchromatin domains can occupy close by regions along the DNA without clear boundary element stopping them from invading into each other. An example would be the region around the boundary of telomeres.

\begin{figure}[btp]
\includegraphics{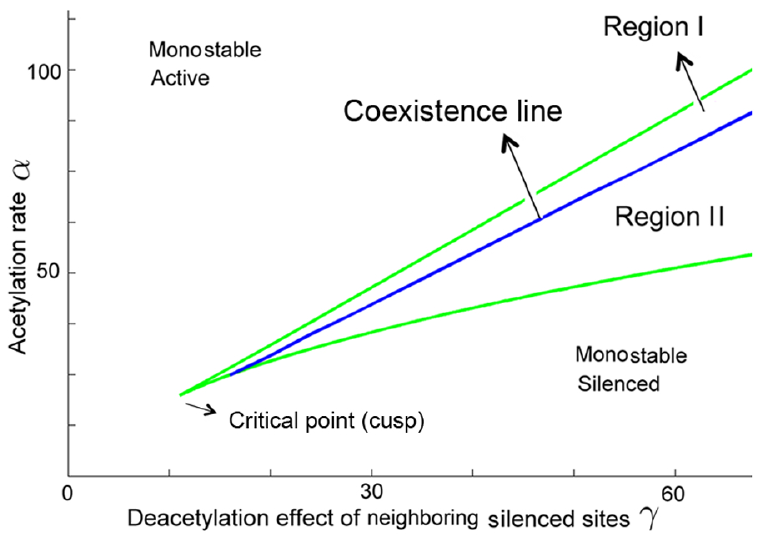}
\caption{Coexistence line subdividing the bistable region. $\rho=30$.} 
\label{fig:alpha_gamma_zero}
\end{figure}

It turns out that in the bifurcation diagram in the $\alpha$ - $\gamma$ plane (constant $\rho$), the condition for the possibility of coexistence of domains define a line where the velocity of the domain boundary is zero (see \ref{APDNU}). We will call this subspace of the parameter space the \textit{coexistence} line, as shown in figure \ref{fig:alpha_gamma_zero}. This line starts from the cusp and divides the bistable regime into two sections. In the lower part, close to the monostable silenced regime, the silenced state is more stable than the active one. The opposite happens in the upper section. In summary, in region I or II of figure \ref{fig:alpha_gamma_zero}, the front between two domains is unstable and moves in the direction of the favorite state; the zero velocity condition defines the coexistence line.

In the above discussion on coexistence of different domains, we considered a continuum system. One might wonder how our results would change if we had, instead, studied a discrete lattice model. To get insight into this, we simulated the stochastic system. As one may have expected, in the discrete version, the coexistence line broadens into a band of propagation failure \cite{sedighi2007epigenetic,keener1987propagation}. In the stochastic version of the model, within this band, the boundary seems to fluctuate without any noticeable drift. In addition, even for very large values of the parameters ($\alpha,\gamma$ and $\rho$), the time scale of fluctuation in the boundary position is quite slow. One of our future plans is to have a theoretical estimate on the relation between boundary fluctuation and the parameters of the system.

\section{Results and discussion}

\subsection{Location of the real system on the bifurcation diagram}
We would like to examine whether the biological model of stepwise spreading of silencing fits into our mathematical description. First, we will assume all the parameters are constant. In Region I of figure \ref{fig:alpha_gamma_zero}, we do not expect the silenced domain to spread from the nucleation center. Instead, this domain should be localized around the nucleation center.  In contrast, in Region II, the silenced domain spreads out from the nucleation center. Although this behavior is similar to the stepwise spreading model, it requires an explicit boundary element to stop it from taking over the whole active domain.

In the telomeric regions of DNA, there does not seem to be an explicit boundary element stopping the spread of silenced domain. For example, by over expressing the SIR proteins, the silenced domain invades into the active one to some extent and then stops again \cite{katan2005heterochromatin}. This implies that the boundary between the two domains is, in principle, dynamic.  At first glance, the stable dynamic boundary between the domains suggests that the system is actually on the coexistence line in figure \ref{fig:alpha_gamma_zero}.   

Assuming the system is on the coexistence line raises two concerns. The first one is that being on this line requires fine tuning of the parameter. The other issue is that if one of the parameters changes, e.g. $\rho$ increases because of over-expression of SIR proteins, the system moves away from the coexistence line. This will cause one domain to invade the other one. However, in reality, this invasion happens only to certain extent and the boundary stabilizes at a new place. So far, our mathematical description does not seem to capture this behavior. 

In the above discussion, we assumed that all the parameters are constant. In particular, the available ambient concentrations of Sir proteins, reflected in $\rho$, was held constant. It turns out that by relaxing this assumption, not only our mathematical description explains the stability of the boundary, but also leads to some interesting predictions. The details are presented in the following part.

\subsection{Consequences of limited supply of Sir proteins}
We can consider the case where the total number of SIR complexes, which is the sum of the proteins in the ambient  and the ones bound to the nucleosomes, is fixed. In other words, there is a limited supply of SIR complexes: 
\begin{equation}\label{eq:limited_supply}
  \rho \ v + \sum_i S_i = S_{tot} = \mbox{constant}  .
\end{equation} 
Here, $v$ is proportional to the volume of the system. This equation means that whenever a SIR complex gets bound to the nucleosome, the ambient concentration of available complexes drops. Therefore, $\rho$ is a self-adjusting parameter, as opposed to being constant. We will see that there will be two implications from this assumption: boundary stabilization and coupling of different silenced regions on the genome. Before going forward, let us look at the bifurcation diagram from another angle.

\begin{figure}[btp]
\includegraphics{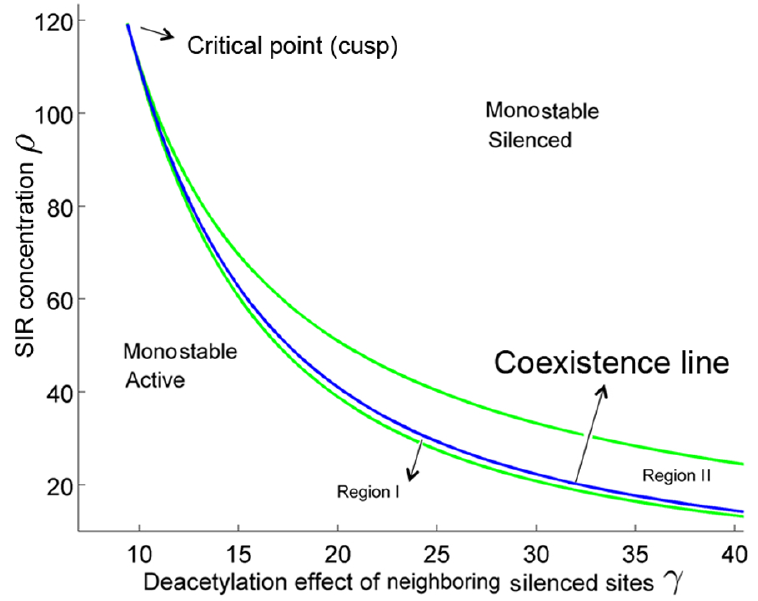}
\caption{The bifurcation diagram in the  $\rho$ - $\gamma$ plane (constant $\alpha$) with $\alpha=60$. The correspondence between different regions of this graph and Figure \ref{fig:alpha_gamma_zero} is shown on the picture.}
\label{fig:jap_1_bifurcation_diagram}
\end{figure}

As we mentioned, the bifurcation diagram is a surface in the three dimensional space formed by $\alpha$, $\gamma$ and $\rho$ axis. So far we have chosen to look at the intersection of this surface with the constant $\rho$ plane (formed by $\alpha$ - $\gamma$ axis). For the following discussion, we change this choice and switch to constant $\alpha$ plane. The diagram, which can be sketched using equations \ref{bif_sol_5} and \ref{bif_sol_6}, is shown in figure \ref{fig:jap_1_bifurcation_diagram}. 

\subsection{Boundary stabilization without requirement for fine-tuning}
Let us now consider a case with a single domain boundary between the silenced and the active region. Consider a system located in Region II of figure \ref{fig:jap_1_bifurcation_diagram} and assume there exist a small silenced domain or silencing has been initiated from a nucleation center. Being the favorite solution in Region II, this silenced domain invades into the active one. However, as silencing is spreading and SIR complexes get bound to the chromosome, the ambient SIR proteins decreases, namely, $\rho$ drops. This means, on figure \ref{fig:jap_1_bifurcation_diagram}, the system moves vertically downward and approaches the coexistence line. In this way, the system automatically goes on the coexistence line and the two silenced and active domains will have a stable boundary between them.

\begin{figure}[btp]
\includegraphics{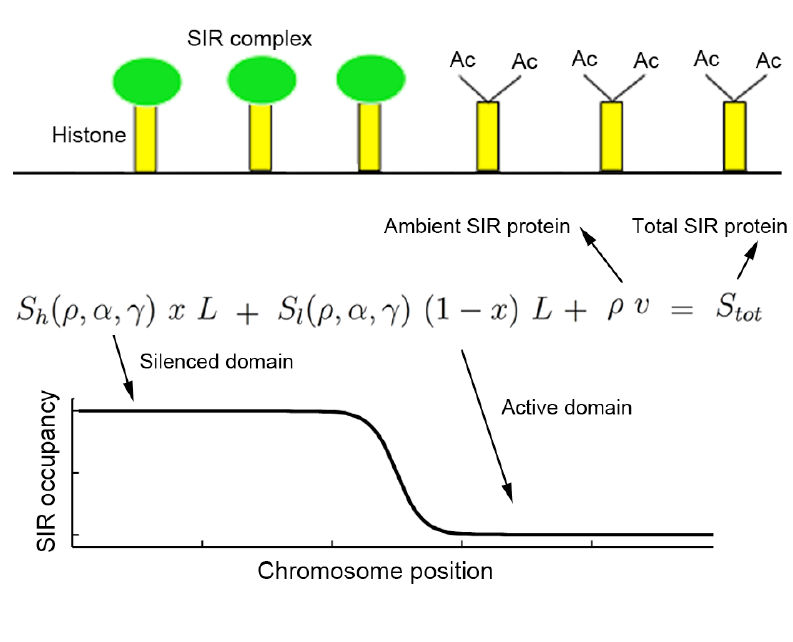}
\caption{A silencing system with a silenced and an active domains in stable coexistence. Size of the chromosomal locus is denoted by $L$, the surrounding volume by $v$ and the total number of SIR proteins by $S_{tot}$. The self-adjusting parameters include the fraction of the system in the silenced state, $x$, and the density of ambient SIR proteins, $\rho$.} 
\label{fig:single_domain_stable}
\end{figure}

The same would have happened if we had started with a system in Region I with some sites in the silenced domain. This time, the silenced domain would shrink and the system moves upward in figure \ref{fig:jap_1_bifurcation_diagram} until it reaches the coexistence line. In the sense of the above discussion, the constraint given by equation \ref{eq:limited_supply} acts as a negative feedback on the perturbation to the system. Figure \ref{fig:single_domain_stable} shows a region with a single domain boundary between the two epigenetic states. In this case, the  constraint given by equation \ref{eq:limited_supply} can be represented by the equation shown in the figure. In \ref{APSAR} we demonstrate how one can determine the location of a system with parameters $S_{tot}$, $v$, $\gamma$ and $L$ (size of the region), in the bifurcation diagram. In other words, it is shown how to calculate the self-adjusting parameters $\rho$ and $x$ (the fraction of the system in the silenced state).  

\subsection{Self-adjusting path in the bifurcation diagram}
The result presented so far on the stabilization of domain boundary is qualitatively similar to the one studied in \cite{sedighi2007epigenetic}. We would like to study the effect of changing the parameters on a bistable silenced system and the implications of titration of SIR proteins in greater detail. Owing to the constraint imposed by equation \ref{eq:limited_supply}, altering any of the parameters produces an adjustment to the abundance of the proteins and creates a non trivial trajectory in the parameter space. As we will see, one  gets qualitatively two different kinds of trajectories as the system traverses the bistable region. Studying the behavior of such trajectories is essential to predict and understand the results of a hysteresis experiment in the silencing system. 

\begin{figure*}[btp]
\includegraphics{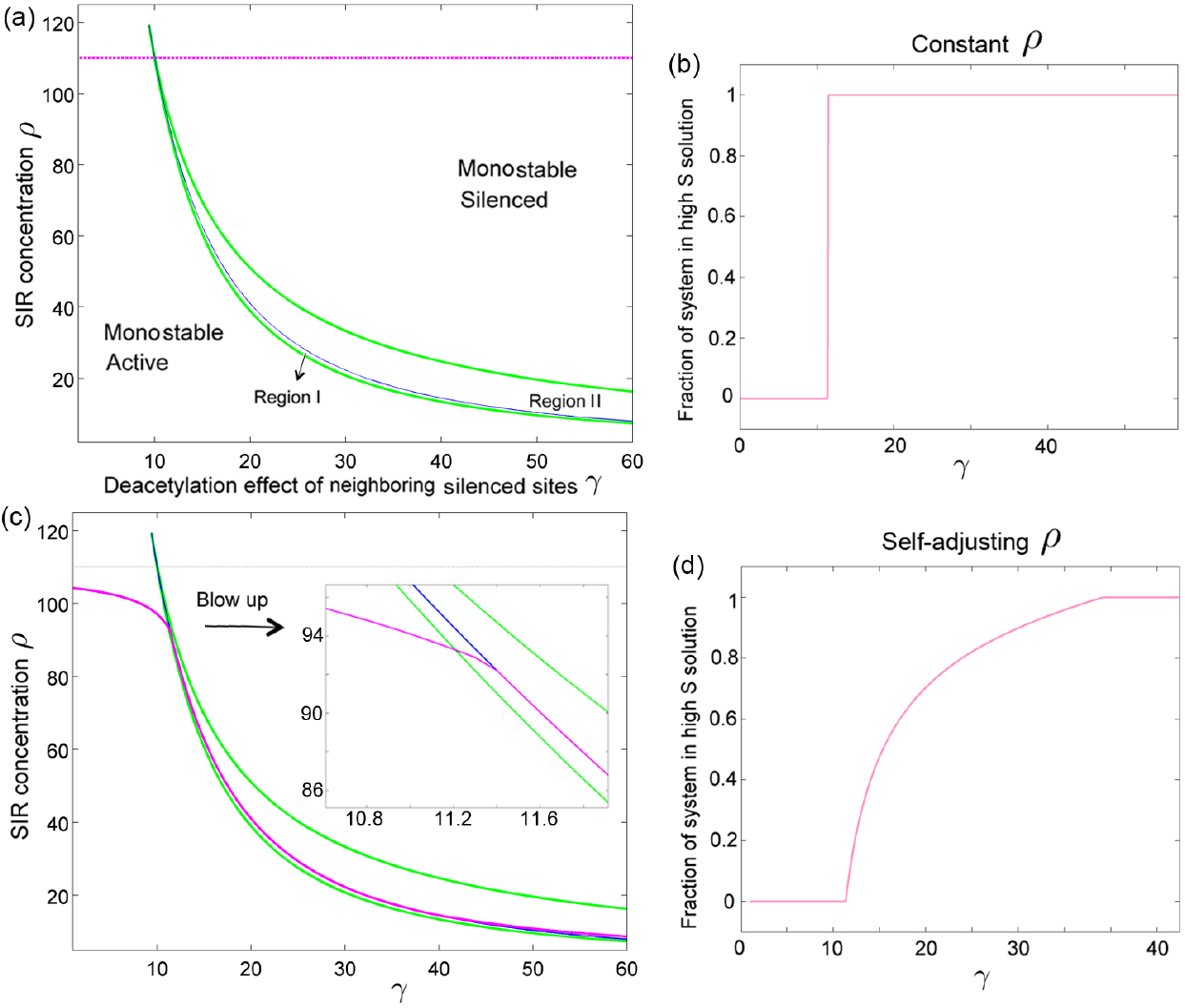}
\caption{The self-adjusting path in the bifurcation diagram as $\gamma$ varies. (a) The path in the case of constant $\rho$ is shown by the magenta line. (b) Fraction of the system in the high $S$ solution for (a). (c) The path in the case of limited supply of $\rho$ (equation \ref{eq:limited_supply}). (d) Fraction of the system in the high $S$ solution for (c). For all panels, $L=200$ sites, $S_{tot}=110$, $\alpha=60$ and $v=1$.}
\label{fig:jap_path}
\end{figure*}

Imagine we have a knob which allows us to play with the value of $\gamma$. In fact, experimentally, such a knob is available. By changing the concentration of nicotinamide (NAM), an inhibitor of Sir2p,  one can effectively modulate $\gamma$ \cite{bitterman2002inhibition}.  We would like to know how a system changes as one varies the value of  $\gamma$. Let us first consider the simple case where $\rho$ is constant, as opposed to being a self-adjusting parameter. Figure \ref{fig:jap_path}(a) shows the path of such a system which is simply a horizontal line (magenta line). Figure \ref{fig:jap_path}(b) shows the fraction of this system in the high $S$ solution. For the case shown here, since the path is close to the cusp point, the size of the Region I and II is  small. However, the shape of the curve in figure \ref{fig:jap_path}(b), up to a shift along the $\gamma$ axis, is independent of how close or far from the cusp the path crosses the bistable regime. As long as we are in the monostable silenced regime or Region II (above the coexistence line) of the bistable regime, the whole system is in the high $S$ domain. As soon as we cross the coexistence line into the Region I and monostable active regime, the whole system will be in the low $S$ domain.  

How about when there is a limited supply of  $\rho$ and the constraint given by equation \ref{eq:limited_supply} is in action? Figure \ref{fig:jap_path}(c) and (d) show an example. In this case, the magenta line in figure \ref{fig:jap_path}(c) and the pink line in  figure \ref{fig:jap_path}(d) are, respectively, the functions $\rho(\alpha,\gamma)$ and $x(\alpha,\gamma)$  satisfying the equation \ref{self_consis_cond}. For very large values of $\gamma$, all the sites are in the high $S$ solution and the system is either in the monostable silenced regime (not shown in the picture) or Region II of the bistable regime. As one decreases $\gamma$, the system hits the coexistence line and the fraction $x$ drops to values lower than 1. Eventually, the silenced domain shrinks to zero and the system enters 
the Region I of the bistable regime and then the monostable active regime.

\begin{figure*}[btp]
\includegraphics{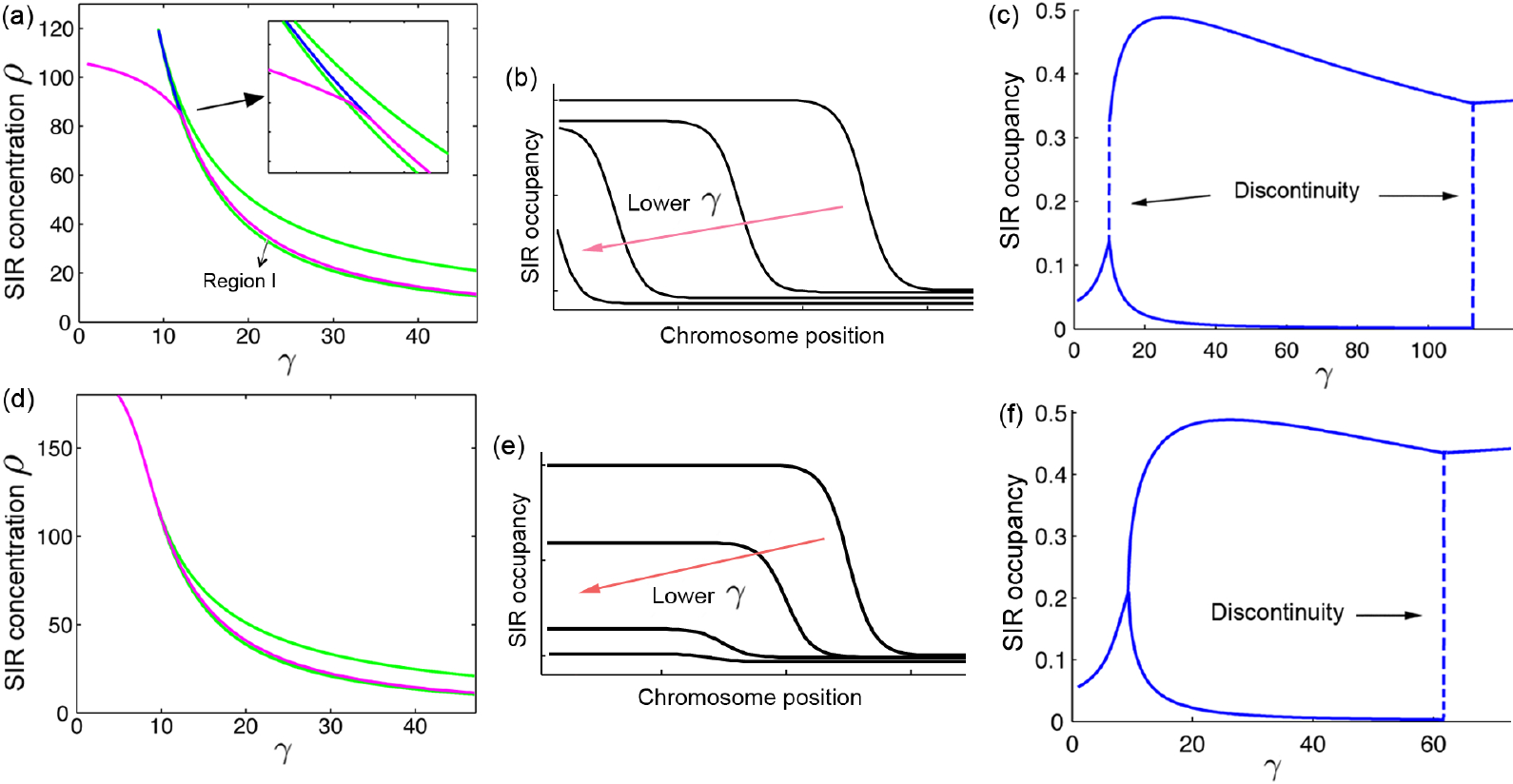}
\caption{Two different possible self-adjusting path in the bifurcation diagram as $\gamma$ varies. (a-c) The case where the size of silencing domain shrinks to zero as $\gamma$ decreases. In this case, the path shown in panel (a) first enters the Region I of the bistable regime before moving on to the monostable active region. (c) The silencing solutions along the self-adjusting path. As the path crosses the bifurcation diagram to enter the monostable active region, one of the two stable $S$ solutions disappears and only one solution remains. There is always a discontinuity between the two stable solutions in the bistable regime. 
(d-f) The case where the bistability is lost before the size of the silencing domain shrinks to zero. (f) As the path crosses the bifurcation diagram through the cusp point to enter the monostable active region, the two stable solutions merge in a continuous manner and form the  single monostable solution. For all panels, $L=500$, $\alpha=60$ and $v=1$. In panels (a-c), $S_{tot}=110$ and in panels (d-f), $S_{tot}=180$. }
\label{fig:two_pathese}
\end{figure*}

In the case of limited supply of SIR proteins, the scenario shown in figures \ref{fig:jap_path}(c) is not the only possibility and one can be in parameter ranges where something else takes place. This point is illustrated in figure \ref{fig:two_pathese}. Figures \ref{fig:two_pathese}(a) shows the scenario that we already discussed. As shown figure \ref{fig:two_pathese}(b), reducing $\gamma$ causes the silencing domain to shrink to its minimum size. 

The other possibility is that before the size of the silencing domain shrinks to zero, bistability is lost all together. In this case, which happens for higher abundances of SIR proteins, the self-adjusting path passes through the cusp point in the bifurcation diagram (figure \ref{fig:two_pathese}(d)). As shown in figure \ref{fig:two_pathese}(e), at the point where the bistability is lost, the size of the silenced domain is non-zero.

We can consider the two silencing solutions in the bistable regime, $S_l(\rho(\alpha,\gamma),\alpha,\gamma)$ and $S_h(\rho(\alpha,\gamma),\alpha,\gamma)$, as well as the single solution in the monostable regime, $S_m(\rho(\alpha,\gamma),\alpha,\gamma)$, along the self-adjusting path. These functions are shown in figure \ref{fig:two_pathese}(c) and (f) and calculated in \ref{APSAR}.  Note that for the case shown in figure \ref{fig:two_pathese}(f), the two solutions for SIR occupancy in the bistable regime merge continuously at the point where the bistability is lost. If the overall supply of SIR proteins is even higher, we could get to a regime where the system remains silenced all the way through the range of variation of the parameter $\gamma$ (not shown here). 

\subsection{Coupling different regions via ambient SIR concentration}
We want to analyze a situation which is inspired by our model system, budding yeast. Each of the 16 chromosomes in a haploid yeast has 2 telomeric regions. In addition, there are two silenced regions, named \textit{HML} and \textit{HMR}, located on chromosome III. The \textit{HML}/\textit{HMR} loci are relatively small in size ($\sim$ 10 sites). In both \textit{HML}/\textit{HMR} loci and telomeres, silencing is initiated by nucleation centers. One important difference is that \textit{HML}/\textit{HMR} loci are surrounded by boundary elements stopping the silencing domain from spreading  \cite{bi1999uasrpg,donze1999boundaries}. On the other hand, telomeric regions have dynamic boundary between silenced and active domains. 

\begin{figure*}[btp]
\includegraphics{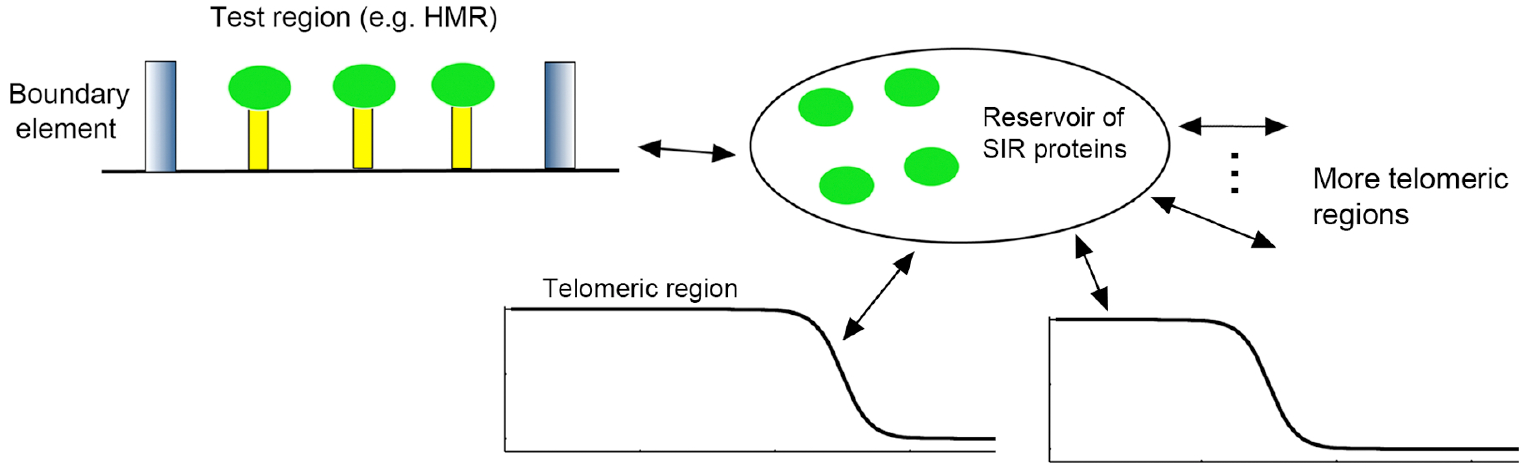}
\caption{Several silencing systems coupled through a shared reservoir of ambient SIR proteins. Perturbing one of the parameters, e.g. $\gamma$, leads to the self-adjustment of the boundary location in regions with free boundary, such as telomeres. Consequently, the concentration of ambient SIR proteins, $\rho$, changes. This, in return, affects the silencing in other regions with explicit boundary elements such as \textit{HMR}.}
\label{fig:reservoir}
\end{figure*}

Our goal is to study the effect of variation in $\gamma$ on this system. Since \textit{HML}/\textit{HMR} loci are small in size, let us ignore their contribution to the constraint imposed by equation \ref{eq:limited_supply}. Telomeric regions have free boundary and from our discussion in the previous section, we know how to determine the self-adjusting path in the bifurcation diagram or equivalently the self-adjusting parameter $\rho(\alpha,\gamma)$. This parameter is the ambient concentration of SIR proteins which is also available to \textit{HML}/\textit{HMR} loci. In other words, \textit{HML}/\textit{HMR} loci read out the value of  $\rho(\alpha,\gamma)$ as it changes due to variation of $\gamma$ and the resulting effect on the state of the telomeric silenced domain (see figure \ref{fig:reservoir}). The possible states for \textit{HML}/\textit{HMR} loci depends on the value of $\rho(\alpha,\gamma), \alpha$ and $\gamma$. In the bistable regime, the two possible silencing levels are denoted by $S_l(\rho(\alpha,\gamma),\alpha,\gamma)$ and $S_h(\rho(\alpha,\gamma),\alpha,\gamma)$. In the monostable regime, there is only one possible solution, $S_m(\rho(\alpha,\gamma),\alpha,\gamma)$. These functions are calculated in \ref{APDUS}. Figure \ref{fig:hysteresis} shows two examples of these functions corresponding to two different parameter regimes described in figure \ref{fig:two_pathese}.

\subsection{Hysteresis at \textit{HML} and \textit{HMR} loci}
\begin{figure*}[btp] 
\includegraphics{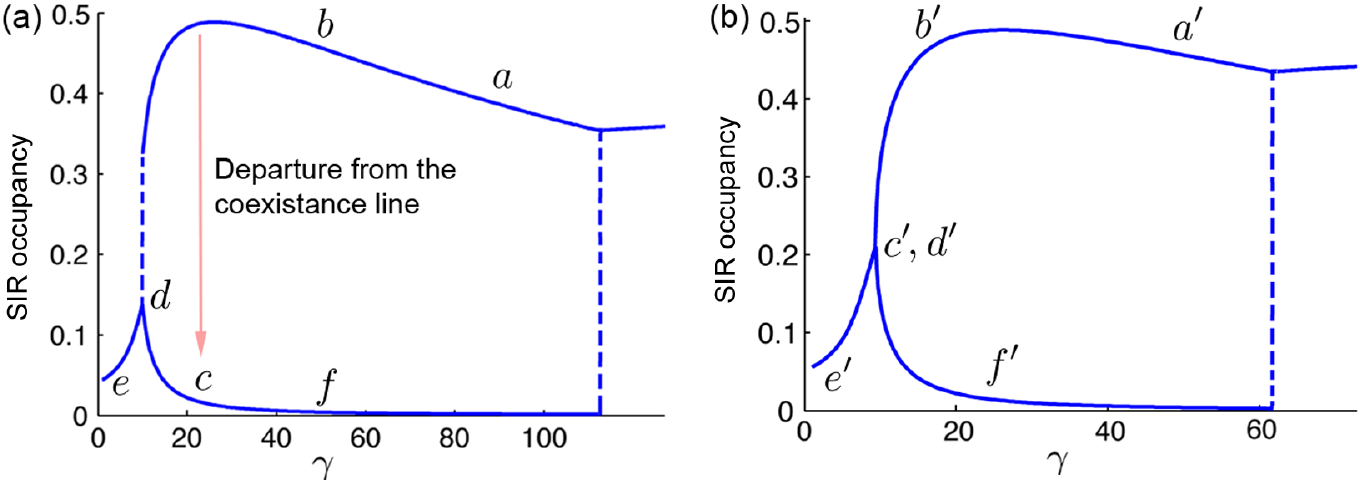} 
\caption{Hysteresis effect in the silencing level at \textit{HML}/\textit{HMR}. (a) Curves represent the silencing solutions along the self-adjusting path shown in figure \ref{fig:two_pathese}(a). Silencing level at \textit{HML}/\textit{HMR} is always located on the blue curves. As $\gamma$ is decreased, the silencing at \textit{HML}/\textit{HMR} follows the points $a\rightarrow b \rightarrow \mbox{(with a jump) } c \rightarrow d \rightarrow e$. Subsequent increase in $\gamma$ changes the silencing value as $e\rightarrow d \rightarrow c \rightarrow f$.  (b) similar to (a) but for the path in figure \ref{fig:two_pathese}(d). Note that the transition between points $b'$ and $(c',d')$ is continuous. 
Moreover, starting from point $e'$, increase in $\gamma$ can take the system to either the lower (point $f'$) or upper (point $b'$) branch.}
\label{fig:hysteresis}
\end{figure*}

Let us start with the following initial condition. The system is initially on the coexistence line as in figure \ref{fig:two_pathese}(a). The silenced domain at telomeric regions coexist with the active domain with a free boundary separating them. Both \textit{HML} and \textit{HMR} loci are in the silenced state, i.e. $S_h(\rho(\alpha,\gamma),\alpha,\gamma)$. Point $a$ in figure \ref{fig:hysteresis}(a) corresponds to the the value of $S_h(\rho(\alpha,\gamma),\alpha,\gamma)$ at this initial state. 

As we decrease $\gamma$, to stay on the coexistence line, $\rho(\alpha,\gamma)$ increases. The new value of $S_h(\rho(\alpha,\gamma),\alpha,\gamma)$ can be obtained as shown in \ref{APSAR}. Point $b$ in figure \ref{fig:hysteresis}(a) corresponds to a value of $\gamma$ for which the system is still on the coexistence line. Note that, as long as the system is on the coexistence line, the change in $S_h$ is continuous. 

An interesting thing happens when the path of the system exit the coexistence line and enters Region I of the bistable regime. By this time, the silencing on the telomeric regions has shrunken to zero. Now, the $S_l(\rho(\alpha,\gamma),\alpha,\gamma)$ solution is more favorite than $S_h(\rho(\alpha,\gamma),\alpha,\gamma)$. Therefore, the state of \textit{HML}/\textit{HMR} loci would change to the lower, more active solution. This transition is shown in figure \ref{fig:hysteresis}(a) by the pink arrow. The new silencing state of the \textit{HML}/\textit{HMR} loci is represented by point $c$ in the figure. As we keep decreasing $\gamma$, the $S_l(\rho(\alpha,\gamma),\alpha,\gamma)$ solution decreases as well. Eventually, the system crosses the bifurcation line (point $d$) and goes into the active monostable regime (point $e$).

What happens if we increase $\gamma$? From point $e$ to $d$ to $c$ in figure \ref{fig:hysteresis}, the state of \textit{HML} and \textit{HMR} loci goes back on the same path as before.  However, at point $c$, where the system hits the coexistence line, the level of silencing at \textit{HML}/\textit{HMR} loci takes a new path. Previously, when we approached this transition point from right, \textit{HML}/\textit{HMR} loci were in the $S_h(\rho(\alpha,\gamma),\alpha,\gamma)$ solution, whereas this time, they are in the $S_l(\rho(\alpha,\gamma),\alpha,\gamma)$ solution. If we increase $\gamma$, the state of these loci will stay on the lower branch and move towards point $f$. One counterintuitive behavior in the above discussion is that, at point $f$ compared to point $c$, although $\gamma$ is higher, the silencing has reduced. The same was also true between points $c$ and $d$:
\begin{equation} \label{uncounter-intuitive_hyst}
S_l^{^f}(\rho(\alpha,\gamma),\alpha,\gamma) < S_l^{^c}(\rho(\alpha,\gamma),\alpha,\gamma)< S_l^{^d}(\rho(\alpha,\gamma),\alpha,\gamma) . \end{equation} 
Intuitively, the decrease in the silencing level at the \textit{HML}/\textit{HMR} loci, despite the increase in the Sir2p activity can be understood as follow: the increase in the Sir2p activity leads to the extension of silenced region at the telomeric regions. This extension depletes the ambient silencing proteins, which results in the reduction of the silencing level at the \textit{HML}/\textit{HMR} loci.

Let us now examine the parameter regime where the self-adjusting path is described by figure \ref{fig:two_pathese}(d), the above picture would get modified. The silencing solutions ($S_h$, $S_l$ and $S_m$)  for this case are depicted in figure \ref{fig:hysteresis}(b). Again, assume that we start at point $a'$ in figure \ref{fig:hysteresis}(b) where the silenced domain at telomeric regions coexist with the active domain with a free boundary separating them and both \textit{HML} and \textit{HMR} loci are in the silenced state, i.e. $S_h(\rho(\alpha,\gamma),\alpha,\gamma)$.  When we decrease $\gamma$, the silencing solution at \textit{HML}/\textit{HMR} loci moves to point $b'$ in figure \ref{fig:hysteresis}(b). As the path of the system exit the coexistence line and the bistable regime through the cusp point (see figure \ref{fig:two_pathese}(d)), the silencing solution at \textit{HML}/\textit{HMR} loci changes smoothly and reaches the point denoted by $(c',d')$ in figure \ref{fig:hysteresis}(b). Note that, in contrast to the case shown in figure \ref{fig:hysteresis}(a), there is no jump in the  silencing level at \textit{HML}/\textit{HMR} loci. Further decrease in $\gamma$, will eventually move the silencing level towards the monostable regime (point $e'$).

As we start to increase $\gamma$, the silencing level at \textit{HML}/\textit{HMR} loci goes  from point $e'$ back to point $(c',d')$. At this time, with further increase of $\gamma$, the self-adjusting path in figure \ref{fig:two_pathese}(d) crosses the cusp and moves along the coexistence line. As we mentioned before, the two stable solutions in the bistable regime are equally stable when the system is on the coexistence line. As a result, and in contrast to the case described in figure \ref{fig:hysteresis}(a), the silencing level at \textit{HML}/\textit{HMR} loci does not necessarily follow the active branch (lower branch) towards the point $f'$ in figure \ref{fig:hysteresis}(b). Instead, it can go back from the upper branch towards the point $b'$ with equal probability. In other words, by sweeping $\gamma$ up and down, one does not necessarily obtain the familiar hysteresis loop.

The above discussion was based on a deterministic description of the system. However, that discussion still guides our conclusions for the behavior of a real system with stochastic effects. If we lower Sir2 activity to go to the monostable active region, and then slowly bring the activity up, at some point, we come to the boundary of the bistable region. In case the system is passing through the cusp, one expects that the population breaks into two subpopulation, one following the upper branch, while the other takes the alternative lower branch. As long as the relative weights of the two subpopulation between these two branches remain different for different initial conditions, we still see a hysteresis effect.

\section{Conclusion and outlook}
In this paper, we concern ourselves with hysteresis effect in a model of chromatin silencing in budding yeast. We compute the bifurcation diagram of the system and analyze the result of sweeping up and down one of the parameter, namely, the catalytic activity of Sir2. In particular, we consider the case where there are limited supply of SIR proteins and show that the resulting depletion effects could alter our usual view of hysteresis. Two most interesting observations are the followings.

The first is that, as the Sir2 activity (represented by $\gamma$) is reduced, it is possible to go from the silenced branch of the system to the active branch without a jump. The other counterintuitive observation is that, along the active branch in the bistable regime, the silencing decreases as the Sir2 catalytic activity increases. Both of these observations lead to qualitative predictions that could be compared with experimental results. Features like this are characteristic properties of switches based on chromatin modification where the extent of the silenced domain is a self-adjusting variable and stands in contrast with more familiar examples of switches based on transcriptional regulatory feedback. 

One available set of tools from biomolecular chemistry is a large number of drugs than can alter the activity of histone modifying enzymes \cite{balasubramanyam2003small,grubisha2005small,mai2006small,dekker2009histone}. In fact, the catalytic activity of some of the relevant proteins in the yeast silencing system, such as Sir2 and Sas2, could be affected by such chemical drugs. Of special interest are HDAC  inhibitors of Sir2 like nicotinamide (NAM)  \cite{bitterman2002inhibition} and splitomycin  \cite{posakony2004identification}. By monitoring the presence and absence of long term epigenetic memory as we sweep up and down in the parameter space using such inhibitors, we should be able to directly test the predictions made in this study. 

With available drugs, it is possible to completely remove the Sir2p activity \cite{posakony2004identification}. In other words, one can experimentally achieve the  values of $\gamma$ small enough for the system to be in the monostable active region. On the other hand, in the wild-type system, $\gamma$ is high enough such that the system  falls in the bistable region. Therefore, the range of experimentally accessible values for $\gamma$ is broad enough to allow testing the predictions of our model. 

The implications of having limited supply of silencing proteins in cells, as discussed here in the context of budding yeast, can be present in some other chromatin silencing systems. The crucial ingredient is that the silencing proteins need to be present throughout the silencing domain, as opposed to be localized in a small region and affect histones at far away distances. In mammalian systems involving Polycomb group genes, the above requirement is believed to be satisfied \cite{boyer2006polycomb,lee2006control,bracken2006genome,squazzo2006suz12}, although it has been challenged in the context of Polycomb group genes in flies \cite{schwartz2007polycomb}.

\section{Acknowledgements}
We thank Vijaylakshmi Nagaraj and Mohammad Sedighi for many useful discussions.  AD was supported by NSF PHY11-25915 while AMS acknowledges support of  National Human Genome Research Institute grant R01HG003470. AMS thanks the Kavli Institute for Theoretical Physics for supporting his visit during which part of this project was completed.

\appendix  
\setcounter{section}{0}
\section{Derivation of uniform solutions} \label{APDUS}
\begin{figure*}[btp]
\includegraphics{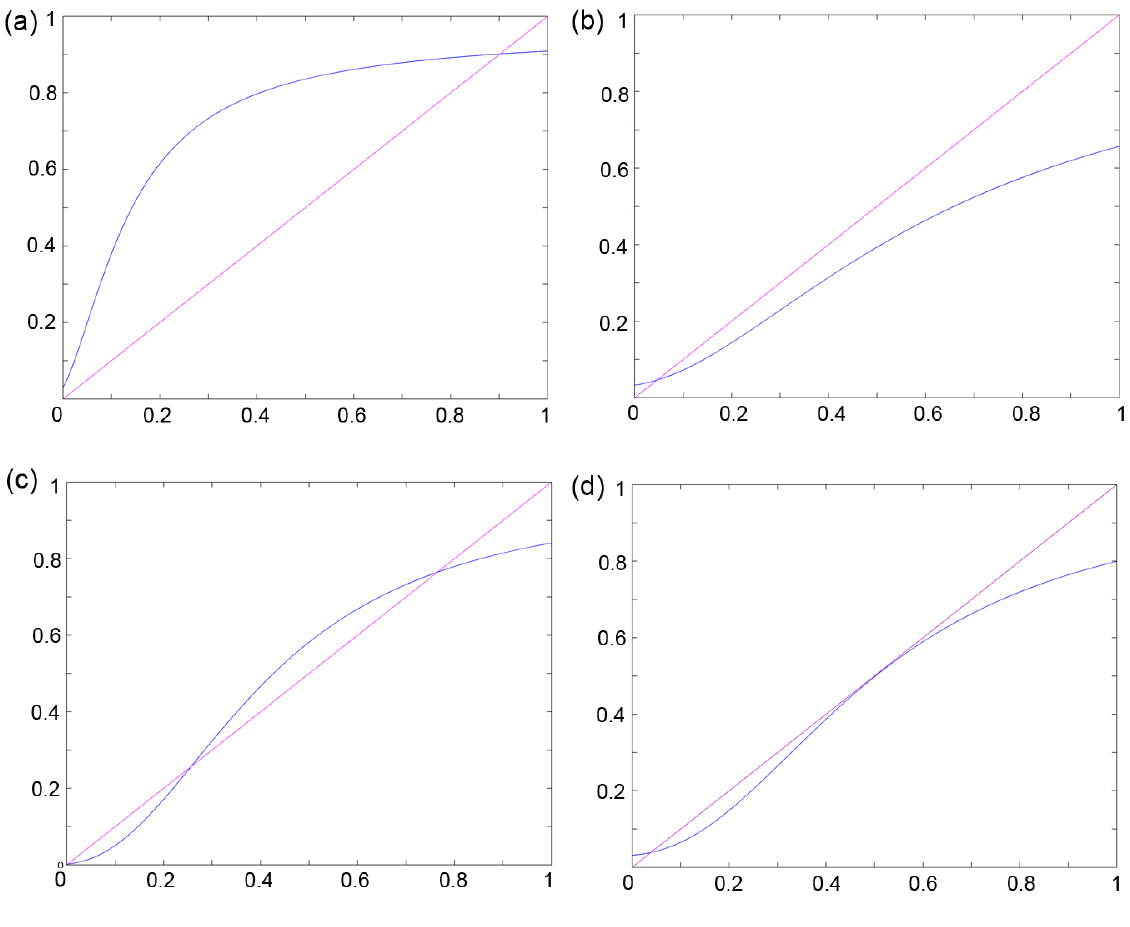} 
\caption{The intersections of nullcline curves. Graphs shows the left hand side (magenta) and the right hand side (blue) of equation \ref{s_sol_0}, for different sets of parameters. (a) Monostable silenced regime. $\alpha=30$, $\rho=30$ and $\gamma=40$. (b) Monostable active regime. $\alpha=115$, $\rho=30$ and $\gamma=40$.  (c) Bistable regime with two stable solutions and one unstable solution. $\alpha=85$, $\rho=30$ and $\gamma=40$. (c) System on the bifurcation line between the bistable regime and the monostable active regime. $\alpha=97$, $\rho=30$ and $\gamma=40$. } 
\label{fig:sgraph}
\end{figure*}

By eliminating the variables in the equation \ref{silencing_uniform}, we find: 
\begin{equation}
E  =  \frac{\eta S}{\rho} ; \qquad A  =  \frac{2 \alpha \eta S}{\rho \ (\lambda+\gamma S)}  ; \qquad
D  =  \frac{\alpha^2  \eta S}{\rho \ (\lambda+\gamma S)^2}    .
\end{equation}
Let us also rescale the parameters by $\lambda$ as follow: 
\begin{equation}\label{}
 \rho'=\frac{ \rho}{\eta};  \qquad \alpha'=\frac{ \alpha}{\lambda}; \qquad  \gamma'=\frac{\gamma}{\eta} .
\end{equation}
From now on, we drop the primes on the rescaled parameters. Since the sum of the probabilities has to be 1, we have: 
\begin{equation*}
S \left(1 + \frac{1}{\rho} +  \frac{2  \hspace{1mm} \alpha}{\rho \ (1+\gamma S)}  +  \frac{\alpha^2}{\rho \ (1+\gamma S)^2}  \right) = 1\nonumber   ; 
\end{equation*}
which we can rewrite as:
\begin{equation}\label{s_sol_0}
S =\frac{\rho \ (1+\gamma S)^2}{\left[ (1+\rho) \ (1+\gamma S)^2 +  2 \alpha  (1+\gamma S) +  \alpha^2   \right] } . \\
\end{equation}
Figure \ref{fig:sgraph} shows the graph of left hand and right hand side of the above equation as a function of $S$ for a few different combination of the parameters. The above equation is of degree 3 and can have at most three real roots. For certain sets of parameters, referred to as the monostable regime, there is only one real solution (figure \ref{fig:sgraph}(a) and (b)). For relatively small values of $\alpha$ (or high values of $\gamma$ or $\rho$), this solution happens at high $S$ (figure \ref{fig:sgraph}(a)), whereas for relatively high values of $\alpha$, the solution is at low $S$ (figure \ref{fig:sgraph}(b)). There is also a regime of parameters where there are three real solutions (figure \ref{fig:sgraph}(c)). The middle solution is unstable, whereas the other two solutions at low and high values of $S$ are stable. We will denote these two stable solutions by $S_l$ and $S_h$, respectively. When the parameters allow us to have two stable solutions, we are in the bistable regime. As we play with the parameters, for example by increasing/decreasing $\gamma$, two of the three solutions merge together (figure \ref{fig:sgraph}(d)) and we are on the boundary between the bistable and monostable regime. At this point, the curves are tangent to each other. The critical point or the cusp in the bifurcation diagram is defined as the point where all three solutions merge together. 
 
\section{Derivation of bifurcation diagram} \label{APDBD}
In  \ref{APDUS}, it was shown that by changing parameters continuously, one can switch between the two regimes where the set of equations \ref{silencing_uniform} have one or three real solutions.  At the transition between these two state, the two curves in figure \ref{fig:sgraph} touch each other at a point. This is the point where two of the solutions merge and disappear or a degenerate solution appears and eventually give rise to two solutions, depending on the direction that we change the parameters. At this point, not only the equation \ref{s_sol_0} is satisfied, but also the derivative of both sides with respect to $S$ should be equal. Let us rewrite equation  \ref{s_sol_0} as:
\begin{equation}\label{s_sol_1}
\rho \ (1+\gamma S)^2 = S \left[ (1+\rho) \ (1+\gamma S)^2 +  2 \alpha  (1+\gamma S) +  \alpha^2   \right]   . \\
\end{equation}
Putting the derivative of both sides equal, and using  equation \ref{s_sol_1}, we get:
\begin{equation}\label{s_sol_2}
2 \rho \ \gamma S\ (1+\gamma S)  = \rho \ (1+\gamma S)^2 + S^2 \left[2 \gamma (1+\rho) \ (1+\gamma S) +  2 \alpha \gamma \right]  . 
\end{equation}
This implies for the transition point:
\begin{equation}\label{bif_sol_1}
\alpha =  (1+\gamma S)  \left[\frac{\rho (1-S)}{S}  -  1  - \frac{\rho \ (1+\gamma S) }{2 \gamma \ S^2} \right]  .
\end{equation}
After replacing $\alpha$ in equation \ref{s_sol_1} by the above equation and dividing both sides by $S (1+\gamma S)^2$, we get:
\begin{eqnarray*}\label{bif_sol_2}
 \frac{ \rho(1-S)}{S} &=& \left[\frac{ \rho(1-S)}{S}  -   \frac{\rho \ (1+\gamma S) }{2 \gamma \ S^2} \right]^2  \\  &\Longrightarrow &
 \sqrt{\frac{ \rho (1-S)}{S}} =  \frac{ \rho(1-S)}{S}  -   \frac{\rho \ (1+\gamma S) }{2 \gamma \ S^2}  .
\end{eqnarray*} 
The reason we take the positive root is because $\alpha >  0$; therefore, the term in the bracket in equation \ref{bif_sol_1} is positive.  We can solve the above equation for $\gamma$:
\begin{equation}\label{bif_sol_3}
\gamma = \left(S-2S^2-\sqrt{\frac{4(1-S)S^3}{\rho}}\right)^{-1}  .
\end{equation}
We can replace the above in equation \ref{bif_sol_1} to get:
\begin{equation}\label{bif_sol_4}
\alpha =  \frac{\left(2(1-S)-\sqrt{4\rho^{-1}(1-S)S}\right) \left(\sqrt{\rho S(1-S)} - S  \right) }{S\left(1-2S-\sqrt{4\rho^{-1}(1-S)S}\right)}  .
\end{equation}
In equations \ref{bif_sol_3} and \ref{bif_sol_4}, for each value of $\rho$, $S$ can take any value between 0 and 1 as long as both $\alpha$ and $\gamma$ are positive real numbers.

There is one more case that we did not mention and is not shown in figure \ref{fig:sgraph}. It is possible to have a situation where all three solutions merge together. This case is similar to figure \ref{fig:sgraph}(d), with the difference that the two curves intersect only at one point. To be in such a situation, in addition to equations \ref{s_sol_1} and \ref{s_sol_2}, the second derivative of both sides of the equation \ref{s_sol_1} with respect to $S$ has to be equal. Putting the second derivatives equal, and using equation \ref{s_sol_1} and equation \ref{s_sol_2}, we get:
\begin{equation}\label{bif_sol_7}
S_{C} =  \frac{\gamma \rho -2 \alpha -2(1+\rho) }{3\gamma (1+\rho)}    .
\end{equation}
The subscript $C$ is meant to indicate the critical point or the cusp. Note that the parameters in the above equation have to satisfy equations \ref{s_sol_1} and \ref{s_sol_2} as well.

Equations \ref{bif_sol_3} and \ref{bif_sol_4} are the consequence of the two conditions \ref{s_sol_1} and \ref{s_sol_2} that we have imposed. Instead of solving for 
$\alpha$ and $\gamma$, we could have chosen to solve for $\alpha$ and $\rho$, or $\rho$ and $\gamma$. If we solve for $\rho$ and $\gamma$ in equations \ref{s_sol_1} and \ref{s_sol_2}, we find:
\begin{eqnarray}\label{bif_sol_5}
\gamma & = & \frac{(\alpha-2(1+\alpha S))}{2S} \pm \sqrt{\left[\frac{(\alpha-2(1+\alpha S))}{2S}\right]^2-\frac{(1+\alpha)}{S^2}}\   , \\ \label{bif_sol_6}
\rho & = &  \frac{4(1-S)S^3}{\left(S-2S^2-\gamma^{-1}\right)^2}   ,
\end{eqnarray}
where $\gamma$ in the second equation has to be replaced from the first one.

\section{Derivation of non-uniform solutions} \label{APDNU}
Given that we are dealing with a spatially extended system, in addition to uniform solutions for equation \ref{silencing_uniform}, we would like to explore the possibility of having non-uniform spatial solutions for the parameter sets located in the bistable regime. 
For a system with $N$ nucleosomes, there are $4^N$ possible distinct states. We cannot  directly solve the time-independent solutions of the stochastic system given by the set of equations \ref{silencing_discrete}. Therefore, we will resort to the continuum limit approximation. In this limit, set of equations \ref{silencing_discrete} become:
\begin{eqnarray} \label{silencing_contin}
\frac{\rmd S(x)}{\rmd t} & = & \rho  E(x) - \eta S(x) \ ,  \\
\frac{\rmd E(x)}{\rmd t} & = &   \eta S(x)   -\rho E(x)  -2\alpha  E(x) + \nonumber \\ & & \left(\lambda+\int \Gamma(x-y) S(y) dy\right) A(x)  \ , \nonumber  \\
\frac{\rmd A(x)}{\rmd t} & = & 2\alpha  \  E(x) +2 \left(\lambda+\int \Gamma(x-y) S(y) dy\right) D(x) - \nonumber \\ & &  \left(\alpha+\lambda+\int \Gamma(x-y) S(y) dy\right) A(x) \ , \nonumber \\
\frac{\rmd D(x)}{\rmd t} & = & \alpha  A(x) -2 \left(\lambda+\int \Gamma(x-y) S(y) dy\right) D(x)  . \nonumber 
\end{eqnarray}
We can Taylor expand $S(y)$ around $x$. Since $\Gamma(x-y)$ falls sharply as $|x-y|$ increases, we will only keep up to the third order in the expansion:
\begin{eqnarray}\label{zero_vel_0}
\int \Gamma(x-y) S(y) dy &=& \int \Gamma(x-y)  \Big( S(x)+(y-x)\frac{\rmd S(x)}{\rmd x}+
\nonumber \\ & &
\frac{(y-x)^2}{2}\frac{\rmd^2S(x)}{\rmd x^2} +...\Big) dy  \nonumber \\ & \simeq & \gamma S(x) + \gamma_2 \frac{\rmd^2S(x)}{\rmd x^2}  .
\end{eqnarray}
The second term in the Taylor expansion disappears since $\Gamma(x-y)$ is symmetric. We have also defined:
\begin{equation}\label{zero_vel_1}
  \gamma = \int \Gamma(x-y) dy \ \ \ \& \ \ \   \gamma_2 = \int \Gamma(x-y) \frac{(y-x)^2}{2} dy
\end{equation}
Replacing \ref{zero_vel_0} in the set of equation \ref{silencing_contin} and simplifying the equation we arrive at:
\begin{equation}\label{zero_vel_2}
  \gamma_2 \frac{\rmd^2S(x)}{\rmd x^2}  =  -1 - \gamma S(x) + \alpha \ \frac{S(x)+\sqrt{\rho S(x) (1-S(x))}}{\rho (1-S(x))-S(x)}  .
\end{equation}
If we define:
\begin{equation}\label{zero_vel_pot_def}
   V(S) = +  S + \frac{\gamma}{2} S^2  - \alpha \int^S  \frac{S'+\sqrt{\rho S' (1-S')}}{\rho (1-S')-S'} dS'  .
\end{equation}
then equation \ref{zero_vel_2} can be written in the following form:
\begin{equation}\label{zero_vel_3}
  \gamma_2 \frac{\rmd^2S(x)}{\rmd x^2}  = - \frac{\rmd V(S)}{\rmd S}  .
\end{equation}
The similarity between \ref{zero_vel_3} and the formula for the motion of a particle in a potential field in classical mechanics is clear. In this analogy, $S$ here plays the role of the position of the particle and $x$ plays the role of time. 

\begin{figure*}[btp]
\includegraphics{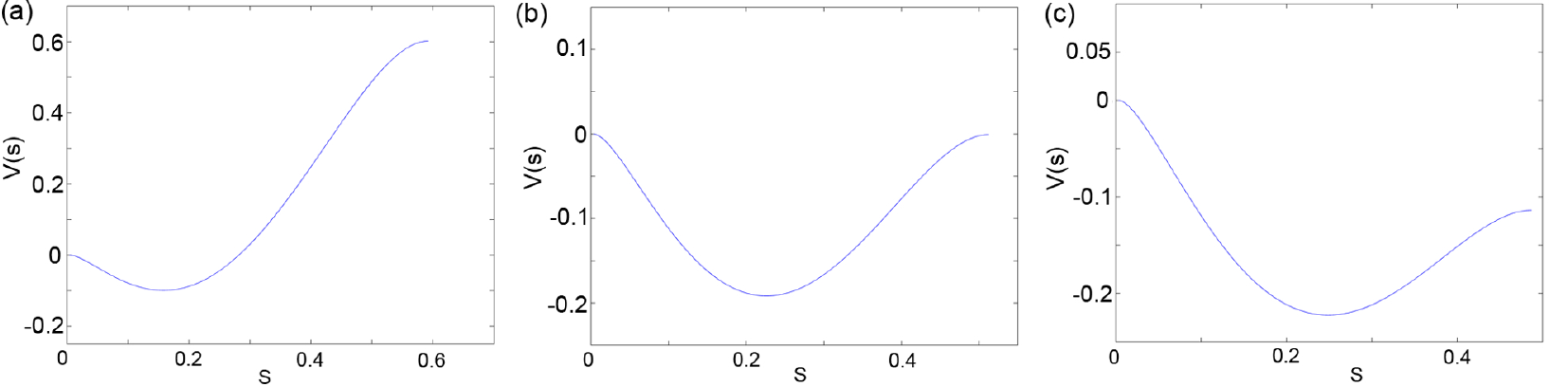}
\caption{Potential $V(s)$  for different combination of parameters in the bistable regime. At each point in the bistable regime, there are two stable solutions for Equation \ref{silencing_contin}, $S_l$ and $S_h$. The graph is only drawn for values of $S$  between these two solutions. (a) $\alpha=90$, $\rho=20$ and $\gamma=55$.  (b) $\alpha=90$, $\rho=20$ and $\gamma=50$. (c) $\alpha=90$, $\rho=20$ and $\gamma=49$. The correspondence with figure \ref{fig:alpha_gamma_zero} is as follow: panel (b) corresponds to the coexistence line; panel (a) is located in Region II; panel (c) is in Region I.} 
\label{fig:bistable_potential}
\end{figure*}

For the two uniform stable solutions of Equations \ref{silencing_contin}, $S_l$ and $S_h$, the right hand side of equation \ref{zero_vel_2} is zero. At those points, the potential $V$, defined in equation \ref{zero_vel_pot_def}, is flat (${\rmd V \over \rmd s}|_{S_l}={\rmd V \over \rmd s}|_{S_h}=0$). Using equation \ref{zero_vel_pot_def}, we can numerically calculate the values of $V$ for the points between the two stable solutions. Figure \ref{fig:bistable_potential} shows the result of numerical integration for different combination of parameters within the bistable regime. 

We are looking for a solution which starts from one of the stable solutions (e.g. $S_l$) and ends in the other solution (e.g. $S_h$).  From our experience in classical mechanics with equations in the form of  \ref{zero_vel_3}, we now that the necessary condition is (Figure \ref{fig:bistable_potential}(b)): 
\begin{equation}\label{equal_pot}
  V(S_l) = V(S_h)  .
\end{equation}
It turns out that in the bifurcation diagram in the $\alpha$ - $\gamma$ plane (constant $\rho$), this condition defines a line that we will call the \textit{coexistence} line. Figure \ref{fig:alpha_gamma_zero} shows this coexistence line.  Note that we can use this potential only to describe the zero-velocity fronts, and the general traveling solution.  

\section{Configuration of a system with limited supply of SIR proteins} \label{APSAR}
If the system is in the bistable regime, there are two possible states, $S_l(\rho,\alpha,\gamma)$ and $S_h(\rho,\alpha,\gamma)$. In the silenced monostable or active monostable region, there is only one possible solution, $S_m(\rho,\alpha,\gamma)$. Note that for fixed $\alpha$ and $\gamma$, these solutions are monotonically increasing function of $\rho$.  

\begin{figure}[btp]
\includegraphics{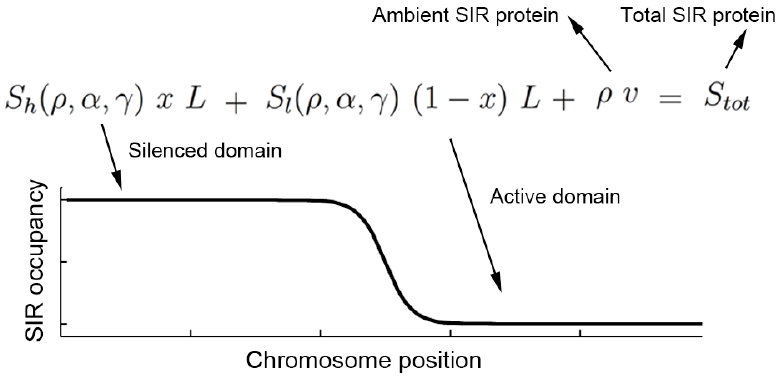}
\caption{Constraint imposed by limited supply of SIR proteins.}
\label{fig:jap_telomere_1}
\end{figure}

Consider a particular value of $\gamma$ and $\alpha$. Assume this value is high enough so that for certain  range of $\rho$ the system is in the bistable regime. Lets consider the following function:
 
\begin{eqnarray} \label{self_consis_rho_state}
&&\phi(\rho,\alpha,\gamma,x)   =   \\ \nonumber \\
&& \left\{ 
\begin{array}{ll} 
\phi_1 = S_m(\rho,\alpha,\gamma) \ L + \rho \ v &  \\  \hspace{1cm} \mbox{\small{if} }  \rho  \mbox{ \small{in active monostable region,}}&  \vspace{.3cm} \\  
\phi_2 = S_l(\rho,\alpha,\gamma) \ L + \rho \ v  & \\  \hspace{1cm} \ \mbox{\small{if} }  \rho  \mbox{ \small{in Region I of bistable regime,}}& \vspace{.3cm} \\  
\phi_3 = S_l(\rho,\alpha,\gamma) \ (1-x) \ L \ +S_h(\rho,\alpha,\gamma) \ x \ L  + \rho \ v &   \\  \hspace{1cm} \mbox{\small{if} }  \rho  \mbox{ \small{on the coexistence line and where}} \ 0  \le x  \le 1 , &  \vspace{.3cm} \\ 
\phi_4 = S_h(\rho,\alpha,\gamma) \ L + \rho \ v  &  \\  \hspace{1cm} \mbox{\small{if} }  \rho  \mbox{ \small{in Region II of bistable regime,}}  \vspace{.3cm} \\ 
\phi_5 = S_m(\rho,\alpha,\gamma) \ L + \rho \ v &   \\  \hspace{1cm} \mbox{\small{if} }  \rho  \mbox{ \small{in silenced monostable region}}.
\end{array} \right.    \nonumber 
\end{eqnarray}   
The variable $x$ represents the fraction of the system in the $S_h$ domain. It can be different from zero or one only for a particular value of $\rho$ for which the system is located on the coexistence line (See figure \ref{fig:jap_telomere_1}). Note that
$\phi$ is a monotonically increasing function of $\rho$ and $x$.

For a fixed value of $\gamma$, $\alpha$ and $S_{tot}$, to determine what the configuration of a system is and where in the bifurcation it is located, one has to find:
\begin{equation}\label{self_consis_cond}
\rho(\alpha,\gamma) \mbox{ and } x(\alpha,\gamma) \mbox{ such that } \phi(\rho,\alpha,\gamma,x) = S_{tot}.   
\end{equation}  

Note that, we should have really written $\rho(\alpha,\gamma,S_{tot},L)$ and $x(\alpha,\gamma,S_{tot},L)$, however, for the sake of brevity, we did not write the last two parameters. How can we calculate $\rho(\alpha,\gamma)$ and $x(\alpha,\gamma)$? Let us consider some particular values of $\rho$ which are of interest. In the active monostable region, the minimum value of $\rho$ is 0 and the maximum value happens when we touch the bifurcation line (green line) in figure \ref{fig:jap_1_bifurcation_diagram} from below. We call this value $\rho_{bu}(\alpha,\gamma)$ ($b$ stands for bifurcation and $u$ for active). Let us refer to the value of $\rho$ on the coexistence line by $\rho_z(\alpha,\gamma)$. As we increase $\rho$, we hit the green line again, this time on the boundary between bistable regime and the silenced monostable one. Let us call this value $\rho_{bs}(\alpha,\gamma)$ ($b$ stands for bifurcation and $s$ for silenced). With this notation, we have: $\rho_{bu} < \rho_{z} < \rho_{bs}$. Using this notation and the definitions given in equation \ref{self_consis_rho_state}, we have: 
\begin{eqnarray}\label{marker-rho}
&&\phi_1(\rho_{bu}(\alpha,\gamma),\alpha,\gamma) < \phi_3(\rho_z(\alpha,\gamma),\alpha,\gamma, x=0) < \nonumber \\ && \phi_3(\rho_z(\alpha,\gamma),\alpha,\gamma, x=1) < \phi_5(\rho_{bs}(\alpha,\gamma),\alpha,\gamma)  . 
\end{eqnarray}    
Also, note that:
\begin{eqnarray}\nonumber
&&\phi_1(\rho_{bu}(\alpha,\gamma),\alpha,\gamma) =\phi_2(\rho_{bu}(\alpha,\gamma),\alpha,\gamma)  \ \mbox{ and } \nonumber \\ &&  \phi_4(\rho_{bs}(\alpha,\gamma),\alpha,\gamma) = \phi_5(\rho_{bs}(\alpha,\gamma),\alpha,\gamma) .
\end{eqnarray}  

The first step in determining the configuration of a system and where in the bifurcation diagram it is located is to compare $S_{tot}$ with the 4 values involved in equation \ref{marker-rho}. If $S_{tot} < \phi_1(\rho_{bu}(\alpha,\gamma),\alpha,\gamma)$, the system is located in the active monostable region. Similarly, if $ S_{tot} > \phi_5(\rho_{bs}(\alpha,\gamma),\alpha,\gamma)$, the system is located in the silenced monostable region. If $\phi_1(\rho_{bu}(\alpha,\gamma),\alpha,\gamma) < S_{tot} < \phi_3(\rho_z(\alpha,\gamma),\alpha,\gamma, x=0)$, the system will be in Region I of figure \ref{fig:jap_1_bifurcation_diagram}. On the other hand, if $\phi_3(\rho_z(\alpha,\gamma),\alpha,\gamma, x=1) < S_{tot} <  \phi_5(\rho_{bs}(\alpha,\gamma),\alpha,\gamma)$, the system will be in Region II.  For each of these regions, one can numerically solve the corresponding $\phi_i$ in equation \ref{self_consis_rho_state} for different values of $\rho$ and find the one that satisfies the constraint given by equation \ref{self_consis_cond}. 

The only remaining case is when
\begin{equation}\nonumber
\phi_3(\rho_z(\alpha,\gamma),\alpha,\gamma, x=0) < S_{tot} < \phi_3(\rho_z(\alpha,\gamma),\alpha,\gamma, x=1) ,
\end{equation}
which corresponds to a system with domains of silenced and active regions coexisting with each other. The fraction of the system in the $S_h$ domain is determined by satisfying:
\begin{eqnarray} \nonumber
S_l(\rho_z(\alpha,\gamma),\alpha,\gamma) \ (1-x) \ L \ + && S_h(\rho_z(\alpha,\gamma),\alpha,\gamma) \ x \ L  +  \nonumber \\ &&  \rho_z(\alpha,\gamma) \ v  =   S_{tot}  ,
\end{eqnarray}  
which implies:
\begin{equation} \label{find_x}
x  = \frac{(S_{tot} -  \rho_z(\alpha,\gamma) \ v - S_l(\rho_z(\alpha,\gamma),\alpha,\gamma) \ L)}{(  S_h(\rho_z(\alpha,\gamma),\alpha,\gamma)\   L -S_l(\rho_z(\alpha,\gamma),\alpha,\gamma) \   L )} .
\end{equation}  
In summary, we showed how to calculate the self-adjusting parameter $\rho$ and the configuration of the corresponding system. 


\end{document}